\documentclass[pra,twocolumn,tightenlines,nofootinbib]{revtex4}
\usepackage{graphicx}
\usepackage{multirow}
\usepackage{amsmath,amsfonts,amssymb}
\usepackage{bm}
\usepackage{dcolumn}
\usepackage{graphicx}

\newcommand{\eref}[1]{Eq.~(\ref{#1})}
\newcommand{\tref}[1]{Table~\ref{#1}}

\begin{document}
\title{Clock-related properties of Lu$^+$}
\author{S.~G.~Porsev$^{1,2}$}
\author{U.~I.~Safronova$^{3}$}
\author{M.~S.~Safronova$^{1,4}$}

\affiliation{
$^1$Department of Physics and Astronomy, University of Delaware, Newark, Delaware 19716, USA\\
$^2$Petersburg Nuclear Physics Institute of NRC ``Kurchatov Institute'', Gatchina, Leningrad District, 188300, Russia,\\
$^3$Physics Department, University of Nevada, Reno, Nevada 89557, USA\\
$^4$Joint Quantum Institute, National Institute of Standards and Technology and the University of Maryland,
College Park, Maryland 20742, USA
}

\date{\today}

\begin{abstract}
Singly-ionized lutetium has a number of fortuitous properties well suited for a design of an optical clock and corresponding applications. In this work, we  study Lu$^+$ properties relevant to a development of the clock using the relativistic high-precision method combining configuration interaction and the linearized coupled-cluster approaches. The systematic effects due to interaction of an external electric-field gradient with the quadrupole moment and the dynamic correction to the blackbody radiation shift are studied and uncertainties are estimated. The value of the $5d6s\,^1\!D_2$  polarizability is predicted.
We also demonstrate that Lu$^+$ is a good candidate to search for variation of the fine-structure constant.
\end{abstract}

\maketitle
\section{Introduction}
\label{Intro}
Further development of frequency standards is important for many applications requiring an improved precision and high stability,
such as searches for the variation of the fundamental constants \cite{SafBudDem18}, tests of the Lorentz invariance \cite{PihGueBai17,ShaOzeSaf18}, dark matter searches~\cite{Arv15,RobBleDai17,KalYu17},  study of many-body physics and quantum simulations \cite{ZhaBisBro14,KolBroBot17},  relativistic geodesy \cite{MehGroLis18}, very long baseline interferometry \cite{NorCle11}, gravitational wave detection \cite{KolPikLan16} and others.  The systematic uncertainties at the $10^{-18}$ level, two orders of magnitude better than for the Cs clock currently defining the SI second~\cite{Bau03}, were recently demonstrated with both neutral atom lattice clock based on the $^1\!S_0 -\, ^3\!P_0^o$ transition in Sr~\cite{NicCamHut15} and a single trapped ion clock based on the electric octupole $^2\!S_{1/2} -\, ^2\!F^o_{7/2}$ transition in $^{171}$Yb$^+$ \cite{HunSanLip16}.

A bottleneck to an improvement of the trapped ion clocks is  the relatively low stability achieved with a single ion. Proposed solutions of this problem include a development of clocks with ion chains ~\cite{KelBurKal19} and large ion crystals \cite{ArnHajPae15,ArnBar16}.
In the recent paper~\cite{KelBurKal19} the authors have demonstrated a possibility to control systematic frequency uncertainties at the 10$^{-19}$ level in linear Coulomb crystals for In$^+$ clock sympathetically cooled with Yb$^+$ ions.

An important problem affecting both the  neutral atom and trapped ion clocks, is the blackbody radiation (BBR) shift \cite{PorDer06BBR}. Small BBR shift at room temperature is a highly desirable feature that simplifies the clock design
removing the requirement to maintain either precise temperature control \cite{NicCamHut15} or cryogenic cooling \cite{UshTakDas15}.

A singly ionized lutetium was suggested as a promising novel clock candidate, having
a number of favorable properties leading to low systematic shifts \cite{Bar15,ArnHajPae15,ArnBar16}.
There are two clock transitions with favorable systematics, the highly forbidden $6s^2\,^1\!S_0 -\, 5d6s\,^3\!D_{1}$ $M1$ transition at 848~nm and the $6s^2\,^1\!S_0 -\, 5d6s\,^3\!D_{2}$ $E2$ transition at 804~nm. A joint experimental
and theoretical investigation of the $6s^2\,^1\!S_0 -\, 5d6s\,^3\!D_{1,2}$ clock transitions in the Lu$^+$ was carried out in Ref.~\cite{PaeArnHaj16}. The dc and ac polarizabilities of the clock states, lifetimes of the low-lying states, hyperfine quenching rate of the $6s6p\,^3\!P_0^o$ state, and other properties were reported.
The BBR frequency shift of the $^1\!S_0 -\, ^3\!D_1$ clock transition was also calculated in Ref.~\cite{KozDzuFla14}.

In 2018, the  differential scalar polarizabilities of these clock transitions were measured at the
wavelength $\lambda = 10.6\, \mu{\rm m}$  in Ref.~\cite{ArnKaeRoy18} to be $\Delta \alpha_0(^3\!D_{1} -\, ^1\!S_0) = 0.059(4)$~a.u.
and $\Delta \alpha_0(^3\!D_{2} -\, ^1\!S_0) = -1.17(9)$~a.u..  From this, Arnold {\it et al}. \cite{ArnKaeRoy18} extracted  the fractional
BBR frequency shift for the $^1\!S_0 -\, ^3\!D_{1}$  transition to be $-1.36(9) \times 10^{-18}$ at
300 K. This shift is the lowest of any established atomic optical clocks. In particular, it is a factor of six smaller than the
fractional BBR shift for the $^1\!S_0 -\, ^3\!P_0^o$ transition in Al$^+$  \cite{CheBreCho17}.

Another important systematic issue, crucial to an operation of ion clocks, is the micromotion-induced shift. It is driven
by the rf-trapping field and leads to an ac Stark shift and a second-order Doppler shift.
If the differential scalar polarizability of the clock transition is negative, there is a trap drive frequency at which the ac Stark and second-order Doppler shifts cancel each other and the micromotion shift vanishes~\cite{BerMilBer98,DubMadZho13}.
A suppression of this effect in a case of ion clock operating with large ion crystals was discussed in Ref.~\cite{ArnHajPae15}.
Thus, the negative sign of the $^1\!S_0 -\, ^3\!D_{2}$ differential polarizability makes this transition a good
candidate for an implementation of the micromotion cancellation scheme.

In a discussion of the  experimental scheme in Ref.~\cite{PaeArnHaj16} it was noted that an optical pumping via the $^3\!P_1^o$ level
leads to an undesired population of the $^1\!D_2$ state. A decay of this state during optical pumping may be significant systematic effect. The $^1\!S_0 -\, ^1\!D_2$ transition can be used for diagnostic measurements and potentially a clock transition \cite{Mur18}.
As a result, it is important to calculate its properties, in particulary the polarizability and the quadrupole shift.

Thus, further investigations of the clock-related properties of Lu$^+$ and corresponding systematic shifts are urgently needed, which is the subject of this work.
In Sec.~\ref{Polar} we study relevant properties of the $5d6s\,^1\!D_2$ state, including $E1$ transition amplitudes and the static polarizability. In Sec.~\ref{El_quad} we discuss the systematic effect caused by the interaction of external electric-field gradient with the quadrupole moment of an atomic state. In Sec.~\ref{BBR}, we calculate dynamic corrections to the BBR shifts of the $6s^2\,^1\!S_0$, $5d6s\,^3\!D_{1,2}$, and  $5d6s\,^{1}\!D_2$ energy levels. Section~\ref{alphas} is devoted to study of sensitivity of Lu$^+$ to variation of the fine-structure constant and Sec.~\ref{Concl} contains concluding remarks. If not specified otherwise, we use atomic units.
\section{Method of calculation and the $5d6s\,^1\!D_2$ polarizability}
\label{Polar}
A detailed description of the $^1\!S_0$ and $^3\!D_{1,2}$ polarizability calculations is given in Ref.~\cite{PaeArnHaj16}.
Here we use the same approach to calculate the static polarizability of the $5d6s\,^1\!D_2$ state of Lu$^+$. We use the high-precision relativistic methods, combining configuration interaction (CI) with the many-body perturbation theory (MBPT) or with the linearized coupled-cluster (all-order) method~\cite{DzuFlaKoz96,SafKozJoh09}.
The energies and wave functions are determined from the time-independent multiparticle Schr\"odinger equation
\begin{equation}
H_{\rm eff}(E_n) |n \rangle = E_n |n \rangle,
\label{Heff}
\end{equation}
with the effective Hamiltonian defined as
\begin{equation}
H_{\rm eff}(E) = H_{\rm FC} + \Sigma(E).
\end{equation}
Here $H_{\rm FC}$ and $\Sigma$ are the Hamiltonian in the frozen core
approximation and the energy-dependent correction, respectively. The latter
takes into account virtual core excitations in the second order of the
perturbation theory (the CI+MBPT method) or in all orders (the CI+all-order method).

The static electric dipole polarizability of the $|0 \rangle$ state is given by
\begin{eqnarray}
\alpha(0) &=& 2\, \sum_k \frac {|\langle k |D_0| 0 \rangle|^2} {E_k-E_0}.
\label{stat_alpha}
\end{eqnarray}
where $D_0$ is the $z$-component of the effective electric dipole operator, including the random-phase approximation (RPA), core-Brueckner ($\sigma$), two-particle (2P), structural radiation (SR), and normalization corrections described in Ref.~\cite{DzuKozPor98}.

The scalar static polarizability $\alpha_0$ can be conventionally separated into three parts:
\begin{equation}
\alpha_0 = \alpha_0^v + \alpha^c + \alpha^{vc}.
\label{alpha}
\end{equation}
Here, $\alpha_0^v$ is the valence polarizability, $\alpha^c$ is the ionic core polarizability, and a small term $\alpha^{vc}$
accounts for possible excitations to the occupied valence shells.
The valence part of the scalar polarizability, $\alpha_0^v$, as well as the tensor polarizability, $\alpha_2$, are calculated by solving inhomogeneous equation in the valence space. We use the Sternheimer~\cite{Ste50} or Dalgarno-Lewis~\cite{DalLew55}
method implemented in the CI+all-order approach~\cite{KozPor99a,PaeArnHaj16}.
The $\alpha^{c}$ and $\alpha^{vc}$ terms are evaluated using the RPA.
The $\alpha^{vc}$ term is calculated as a sum of
contributions from the individual electrons, i.e., $\alpha^{vc}(5d6s)=\alpha^{vc}(5d)+\alpha^{vc}(6s)$.
\begin{table}
\caption{\label{Pol_stat1D2} Contributions to $\alpha_0(0)$ of the $5d6s\,^1\!D_2$ state (in a.u).
The contributions of several lowest-lying intermediate states are listed separately with the corresponding absolute
values of $E1$ reduced MEs given (in a.u.) in column labeled ``$D$''. The theoretical and experimental
\cite{RalKraRea11} transition energies are given (in cm$^{-1}$) in columns $\Delta E_{\rm th}$  and $\Delta E_{\rm expt}$.
We present the contribution of other (not explicitly listed in the table)
intermediate states with fixed total angular momentum $J_n$ in rows labeled ``Other ($J_n = 1,2,3$)''. In rows labeled
``Total ($J_n = 1,2,3$)'' we give the total contribution of {\it all} intermediate states with fixed total angular momentum $J_n$.
In the row ``Total val.'' we present the total value of $\alpha_0^v$.
The dominant contributions to the polarizabilities, listed in columns
$\alpha[\mathrm{A}]$ and $\alpha[\mathrm{B}]$, are calculated with the experimental~\cite{RalKraRea11}
and theoretical energies, respectively.}
\begin{ruledtabular}
\begin{tabular}{lccrdd}
\multicolumn{1}{l}{Contribution} & \multicolumn{1}{c}{$\Delta E_{\rm th}$}
& \multicolumn{1}{c}{$\Delta E_{\rm expt}$} & \multicolumn{1}{c}{$D$} &
\multicolumn{1}{c}{$\alpha[\mathrm{A}]$}
&\multicolumn{1}{c}{$\alpha[\mathrm{B}]$} \\
\hline \\ [-0.3pc]
$^1\!D_2 - 6s6p\;^3\!P_1^o$& 10826 & 11171  & 0.326  &   0.28  &   0.29  \\
$^1\!D_2 - 6s6p\;^1\!P_1^o$& 20615 & 20891  & 0.994  &   1.38  &   1.40   \\
$^1\!D_2 - 5d6p\;^3\!P_1^o$& 33038 & 32717  & 0.144  &   0.02  &   0.02   \\
$^1\!D_2 - 5d6p\;^1\!P_1^o$& 41967 & 41790  & 2.790  &   5.45  &   5.43   \\
 Other ($J_n = 1$)         &       &        &        &   0.64  &   0.64   \\
 Total ($J_n = 1$)         &       &        &        &   7.77  &   7.78   \\[0.3pc]

$^1\!D_2 - 6s6p\;^3\!P_2^o$& 14815 & 15121  & 0.445  &   0.38  &   0.39  \\
$^1\!D_2 - 5d6p\;^3\!F_2^o$& 24230 & 23892  & 2.289  &   6.42  &   6.33  \\
$^1\!D_2 - 5d6p\;^1\!D_2^o$& 28399 & 28126  & 4.017  &  16.79  &  16.63  \\
$^1\!D_2 - 5d6p\;^3\!D_2^o$& 29793 & 29572  & 0.125  &   0.02  &   0.02  \\
$^1\!D_2 - 5d6p\;^3\!P_2^o$& 34204 & 33869  & 1.980  &   3.39  &   3.35  \\
 Other ($J_n = 2$)         &       &        &        &   0.79  &   0.79  \\
 Total ($J_n = 2$)         &       &        &        &  27.78  &  27.50  \\[0.3pc]

$^1\!D_2 - 5d6p\;^3\!F_3^o$& 28010 & 27586  & 1.187  &   1.50  &   1.47  \\
$^1\!D_2 - 5d6p\;^3\!D_3^o$& 31678 & 31401  & 0.352  &   0.12  &   0.11  \\
$^1\!D_2 - 5d6p\;^1\!F_3^o$& 36369 & 35747  & 3.341  &   9.14  &   8.98  \\
 Other ($J_n = 3$)         &       &        &        &   5.75  &   5.75  \\
 Total ($J_n = 3$)         &       &        &        &  16.49  &  16.32  \\[0.3pc]

$\alpha_0^v$               &       &        &        &  52.05  &  51.59  \\
$\alpha^c + \alpha^{vc}$   &       &        &        &   3.74  &   3.74   \\
 Total                     &       &        &        &  55.79  &  55.33   \\
 Recommended               &       &        &        &         &  55.3
\end{tabular}
\end{ruledtabular}
\end{table}

To establish the dominant contributions of the intermediate states to the scalar polarizability, we substitute the electric-dipole matrix elements (MEs) and energies according to the sum-over-states formula,~\eref{stat_alpha}.
Replacing the theoretical energies in the  denominator of~\eref{stat_alpha} for
 dominant contributions by the experimental ones changes the polarizability by less than 1\%.
The contributions of several lowest-lying intermediate states to $\alpha_0(0)$ of the $^1\!D_2$ state, calculated with the experimental
and theoretical energies, are
listed in columns $\alpha[\mathrm{A}]$ and $\alpha[\mathrm{B}]$ in Table~\ref{Pol_stat1D2}. The theoretical and experimental~\cite{RalKraRea11} transition frequencies
are given in columns $\Delta E_{\rm th}$ and $\Delta E_{\rm expt}$ in cm$^{-1}$.
Final absolute values of the corresponding reduced electric-dipole MEs,
calculated using the CI + all-order method and including RPA, $\sigma$, 2P, SR, and normalization corrections are listed in
the column labeled ``$D$'' in a.u..

We also present the contribution of the intermediate states, not explicitly listed in the table, with fixed total angular momentum $J_n=1-3$ in rows labeled ``Other ($J_n = 1,2,3$)''. In rows labeled
``Total ($J_n = 1,2,3$)'' we give the total contribution of {\it all} intermediate states with the fixed total angular momentum $J_n$.
The final value of $\alpha_0^v$ is found as the sum of the values given in these rows.

The contributions from $\alpha^{c}$ and $\alpha^{vc}$ terms are listed together in the respective row.
Taking into account that the main contribution to the $^1\!D_2$ level comes from
the $5d_{5/2} 6s$ configuration (73\%), we determined $\alpha^{vc}$ terms for
the $^1\!D_2$ polarizability as $\alpha^{vc}(5d_{5/2})+\alpha^{vc}(6s)$.
In the row labeled ``Total'' we present the total value of the scalar static $^1\!D_2$ polarizability.
The result obtained with use of theoretical energies, considered as recommended, is given in the row labeled ``Recommended''.

To determine uncertainty of the polarizability we have also calculated its value using two other approximations:
the CI+MBPT+RPA and CI+all-order+RPA. In both cases only RPA corrections were included. The results obtained in the
CI+MBPT+RPA, CI+all-order+RPA, and CI+all-order+AC approximations (where abbreviation ``AC'' means {\it all corrections}, including RPA, $\sigma$, 2P, SR, and normalization) are presented in~\tref{Polariz} in columns (1), (2), and (3), correspondingly.
All calculations are performed with theoretical energies. The uncertainties were estimated as the spread
of the values in columns (1)-(3).

We consider the values obtained in the CI+all-order+AC approximation as the final ones.
A comparison of columns (2) and (3) in~\tref{Polariz} shows that the corrections beyond RPA only slightly
change the value of the $^1\!D_2$ polarizability. Our final result for the $^1\!D_2$ scalar static polarizability
is $\alpha_0(5d6s\,^1\!D_2) = 55.3(1.7)$ a.u..
\begin{table}
\caption{\label{Polariz} The static scalar ($\alpha_0$) and tensor ($\alpha_2$) polarizabilities,
obtained in the CI+MBPT+RPA, CI+all-order+RPA, and CI+all-order+AC approximations,
are presented (in a.u.) in columns (1), (2), and (3), respectively.
 The recommended values are listed in the last column.
The uncertainties are given in parentheses.}
\begin{ruledtabular}
\begin{tabular}{lcccd}
\multicolumn{1}{l}{Polarizability} & \multicolumn{1}{c}{(1)} & \multicolumn{1}{c}{(2)} &
\multicolumn{1}{c}{(3)} & \multicolumn{1}{c}{Recommend.} \\
\hline \\ [-0.3pc]
$\alpha_0(6s^2\;^1\!S_0)^{\rm a}$     & 62.5 &  63.3  &  63.0  &  63.0(0.8)  \\[0.5pc]

$\alpha_0(5d6s\;^1\!D_2)$             & 54.3 &  56.0  &  55.3  &  55.3(1.7)  \\
$\alpha_2(5d6s\;^1\!D_2)$             & 14.5 &  15.2  &  15.7  &  15.7(1.2) \\[0.5pc]

$\alpha_0(^1\!D_2)-\alpha_0(^1\!S_0)$ & -8.2 &  -7.3  &  -7.7  & -7.7(0.9)
\end{tabular}
\end{ruledtabular}
\flushleft $^{\rm a}$These results were obtained in~\cite{PaeArnHaj16}.
\end{table}

\section{Electric quadrupole shift}
\label{El_quad}
The Hamiltonian $H_Q$ describing the interaction of an external electric-field gradient with the quadrupole
moment of an atomic state $|\gamma JIFM \rangle$ (where $J$ is the total electronic angular momentum,
$I$ is the nuclear spin, {\bf F} = {\bf J} + {\bf I} is the total
angular momentum, $M$ is the projection of {\bf F}, and $\gamma$ encapsulates all other electronic quantum numbers)
is given by~\cite{Ita00},
\begin{equation}
H_Q = \sum_{q=-2}^2 (-1)^q \nabla \mathcal{E}^{(2)}_q Q_{q} .
\label{HQ}
\end{equation}
The $q=0$ component of $\nabla \mathcal{E}^{(2)}$ can be written as~\cite{Ram56,Ita00}:
\begin{equation}
\nabla \mathcal{E}^{(2)}_0 = -\frac{1}{2}\,\frac{\partial {\mathcal E}_z}{\partial z}
\end{equation}
and we can estimate the energy shift of the atomic state $|\gamma JIFM \rangle$ as
\begin{equation}
\Delta E \simeq -\frac{1}{2}\, \langle Q_0 \rangle \,\frac{\partial {\mathcal E}_z}{\partial z} ,
\end{equation}
where $\langle Q_0 \rangle \equiv \langle \gamma JIFM |Q_0| \gamma JIFM \rangle$.

Then, the fractional electric quadrupole shift of the clock transition $^3\!D_J -\, ^1\!S_0$ ($J=1,2$) is
\begin{equation}
\frac{\Delta \omega}{\omega} \approx -\frac{1}{2 \omega}\,
\Delta \langle Q_0 \rangle \, \frac{\partial {\mathcal E}_z}{\partial z} ,
\label{del_nu}
\end{equation}
where $\omega$ is the $^3\!D_J -\, ^1\!S_0$ transition frequency,
and $\Delta \langle Q_0 \rangle$ is the difference of the expectation values of $Q_0$ for the upper and lower clock states.
Taking into account that the quadrupole moment of the $^1\!S_0$ state is equal to 0, we have
$\Delta \langle Q_0 \rangle = \langle Q_0(^3\!D_J FM) \rangle$.

The expectation value $\langle Q_0 \rangle$ is given by
\begin{eqnarray}
&&\langle \gamma JIFM |Q_0| \gamma JIFM \rangle = (-1)^{I+J+F} \nonumber \\
&\times&[3M^2 - F(F+1)] \sqrt{\frac{2F+1}{(2F+3)(F+1)F(2F-1)}} \nonumber \\
&\times& \left\{
\begin{array}{ccc}
J & J & 2 \\
F & F & I
\end{array}
\right\}
\langle \gamma J ||Q|| \gamma J \rangle ,
\label{Q0}
\end{eqnarray}
where $\langle \gamma J ||Q|| \gamma J \rangle$ is the reduced ME of the electric quadrupole operator.

In Table~\ref{branching}, we list the diagonal MEs of the magnetic dipole ($M1$) and electric-quadrupole operators and
the electric quadrupole moments $\Theta$, defined as
\begin{equation}
\Theta = 2 \sqrt {\frac{J(2J-1)}{(2J+3)(2J+1)(J+1)}}
\langle \gamma J ||Q|| \gamma J \rangle,
\end{equation}
for the $^3\!D_J$ and $^1\!D_2$ states.  The MEs of the $M1$ operator are given in the Bohr magnetons, $\mu_0= |e|\hbar/(2mc)$
(where $e$ and $m$ are the electron charge and mass, $\hbar$ is the Planck constant, and $c$ is the speed of light).
\begin{table}[!htb]
\caption{\label{branching} The energy levels (in cm$^{-1}$), reduced diagonal MEs of the $M1$ (in $\mu_0$)
and $Q$ (in a.u.) operators, and electric quadrupole moments $\Theta$ (in a.u.) for the $^3\!D_J$ and $^1\!D_2$ states.}
\begin{ruledtabular}
\begin{tabular}{rcrcccc}
\multicolumn{1}{c}{Level} & \multicolumn{1}{c}{Energy} & \multicolumn{2}{c}{} & \multicolumn{1}{c}{Operator}
& \multicolumn{1}{c}{ME}  & \multicolumn{1}{c}{$\Theta$}  \\
\hline \\ [-0.5pc]
$6s5d\ ^3\!D_1$   & 11796 &                  &       & $M1$ & -1.22&            \\
                  &       &                  &       & $E2$ & -3.58&  -1.31     \\[0.4pc]

$6s5d\ ^3\!D_2$   & 12435 &                  &       & $M1$ & -6.33&            \\
                  &       &                  &       & $E2$ & -3.70&  -1.77     \\[0.4pc]

$6s5d\ ^3\!D_3$   & 14199 &                  &       & $M1$ & -12.2&            \\
                  &       &                  &       & $E2$ & -8.16&  -3.98     \\[0.4pc]

$6s5d\ ^1\!D_2$   & 17333 &                  &       & $M1$ & -5.53&            \\
                  &       &                  &       & $E2$ & 0.047&   0.022
\end{tabular}
\end{ruledtabular}
\end{table}

As an example, we estimate the magnitude of the quadrupole shift for the $^3\!D_1 -\, ^1\!S_0$ clock transition, for the bosonic 176 isotope of Lu$^+$ with $I=7$. Since $J=1$ for the $^3\!D_1$ state, the possible values
of $F= 6-8$.

Putting $F=7, M=0$, using for an estimate
$\partial {\mathcal E}_z/{\partial z} = 1\,\, {\rm kV}/{\rm cm}^2 \approx 1.029 \times 10^{-15}$ a.u.~\cite{Mur18}
and  $\langle ^3\!D_1 ||Q|| ^3\!D_1 \rangle \approx -3.58$ a.u., we arrive at
\begin{equation}
\frac{\Delta \nu}{\nu} \approx 6.3 \times 10^{-15} .
\end{equation}
Thus, at typical electric field gradients of $\sim {\rm kV}/{\rm cm}^2$, the quadrupole shifts for the
$^3\!D_J$ states are on the order of a few Hz and should be accounted for. However, it can be suppressed using various  schemes \cite{DubMadBer05,Bar15,ArnBar16}. Taking into account that
\begin{equation}
\sum_{M=-F}^F [3M^2 - F(F+1)] =0 ,
\end{equation}
we obtain from~\eref{Q0},
\begin{equation}
\sum_M \langle \gamma JIFM |Q_0| \gamma JIFM \rangle =0 .
\end{equation}
This is also true for the $H_Q$ operator, \eref{HQ}, as was shown in Ref.~\cite{Ita00}.
So, the quadrupole shift vanishes when averaged over all $M$ states of a given hyperfine state~\cite{DubMadBer05}.

In the specific case of the Lu$^+$ ion, for which $I>J$, averaging over all $F$ states of a fixed $|M| \le I+J$ also
cancels the quadrupole shift~\cite{Bar15}. An advantage of this approach is that, it allows to reduce significantly the number of transitions involved and use magnetically insensitive $M=0$ states.

It is worth noting that the $^1\!D_2$ level has an extremely small quadrupole moment. It is a factor of 2 smaller than
the quadrupole moment for the Yb$^+$ upper, $^2\!F^o_{7/2}$, clock state~\cite{HunOkhLip12}.
In particular, the $F=8,\,M=0$ state  would have a quadrupole shift of just a few mHz for typical experimental conditions,
and averaging schemes for the quadrupole shift cancellation may not even be necessary~\cite{Mur18}.


\begin{table} [t]
\caption{\label{table5} Contributions of the intermediate odd-parity states to the dynamic fractional corrections $\eta_1$,
$\eta_2$, and $\eta = \eta_1 + \eta_2$ of the $6s^2\,^1\!S_0$, $5d6s\,^3\!D_{1,2}$, and $5d6s\,^1\!D_2$ states.
The sums of individual contributions are given in the rows labeled ``Total''.
 The numbers in brackets represent powers of 10.}
\begin{ruledtabular}
\begin{tabular}{lcrrr}
State &\multicolumn{1}{l}{Contrib.}& \multicolumn{1}{c}{$\eta_1$} & \multicolumn{1}{c}{$\eta_2$} & \multicolumn{1}{c}{$\eta$} \\
\hline     \\[-0.6pc]
$6s^2~^1\!S_0$ & $6s6p~^3\!P_1^o$ & 0.000055 & 1.22[-7] & 0.000055 \\
               & $6s6p~^1\!P_1^o$ & 0.000421 & 5.19[-7] & 0.000421 \\
               & $5d6p~^3\!D_1^o$ & 0.000013 & 1.15[-8] & 0.000013 \\
               & $5d6p~^3\!P_1^o$ & 0.000003 & 2.18[-9] & 0.000003 \\
               & $5d6p~^1\!P_1^o$ & 0.000017 & 8.68[-9] & 0.000017 \\
               & Total            & 0.000509 & 0.000001 & 0.000510  \\[0.5pc]
$5d6s~^3\!D_1$ & $6s6p~^3\!P_0^o$ & 0.000372 & 2.80[-6] & 0.000374 \\
               & $5d6p~^3\!P_0^o$ & 0.000040 & 5.00[-8] & 0.000040 \\
               & $6s6p~^3\!P_1^o$ & 0.000223 & 1.44[-6] & 0.000224 \\
               & $5d6p~^3\!D_1^o$ & 0.000093 & 1.48[-7] & 0.000094 \\
               & $5d6p~^3\!P_1^o$ & 0.000049 & 6.03[-8] & 0.000049 \\
               & $6s6p~^3\!P_2^o$ & 0.000009 & 3.71[-8] & 0.000009 \\
               & $5d6p~^3\!F_2^o$ & 0.000185 & 3.85[-7] & 0.000185 \\
               & $5d6p~^1\!D_2^o$ & 0.000048 & 7.71[-8] & 0.000049 \\
               & $5d6p~^3\!D_2^o$ & 0.000074 & 1.08[-7] & 0.000074 \\
               & $5d6p~^3\!P_2^o$ & 0.000003 & 3.67[-9] & 0.000003 \\
               & Total            & 0.001097 & 0.000005 & 0.001102 \\  [0.5pc]    $5d6s~^3\!D_2$ & $6s6p~^3\!P_1^o$ & 0.000403 & 2.81[-6] & 0.000406 \\
               & $6s6p~^1\!P_1^o$ & 0.000015 & 4.03[-8] & 0.000015 \\
               & $5d6p~^3\!D_1^o$ & 0.000042 & 6.89[-8] & 0.000042 \\
               & $6s6p~^3\!P_2^o$ & 0.000071 & 3.21[-7] & 0.000072 \\
               & $5d6p~^3\!F_2^o$ & 0.000105 & 2.29[-7] & 0.000105 \\
               & $5d6p~^1\!D_2^o$ & 0.000000 & 1.70[-0] & 0.000000 \\
               & $5d6p~^3\!D_2^o$ & 0.000066 & 1.00[-7] & 0.000066 \\
               & $5d6p~^3\!F_3^o$ & 0.000156 & 2.67[-7] & 0.000156 \\
               & $5d6p~^3\!D_3^o$ & 0.000061 & 8.32[-8] & 0.000061 \\
               & Total            & 0.000920 & 0.000004 & 0.000924  \\[0.5pc] $5d6s~^1\!D_2$ & $6s6p~^3\!P_1^o$ & 0.000033 & 4.76[-7] & 0.000033 \\
               & $6s6p~^1\!P_1^o$ & 0.000047 & 1.94[-7] & 0.000047 \\
               & $5d6p~^1\!P_1^o$ & 0.000046 & 4.76[-8] & 0.000046 \\
               & $6s6p~^3\!P_2^o$ & 0.000025 & 1.95[-7] & 0.000025 \\
               & $5d6p~^3\!F_2^o$ & 0.000166 & 5.24[-7] & 0.000167 \\
               & $5d6p~^1\!D_2^o$ & 0.000314 & 7.14[-7] & 0.000314 \\
               & $5d6p~^3\!P_2^o$ & 0.000044 & 6.85[-8] & 0.000044 \\
               & $5d6p~^3\!F_3^o$ & 0.000029 & 6.88[-8] & 0.000029 \\
               & $5d6p~^3\!D_3^o$ & 0.000002 & 3.17[-9] & 0.000002 \\
               & $5d6p~^1\!F_3^o$ & 0.000106 & 1.49[-7] & 0.000106 \\
               & Total            & 0.000810 & 0.000002 & 0.000813
\end{tabular}
\end{ruledtabular}
\end{table}
\section {Blackbody radiation shift}
\label{BBR}
The leading contribution to the multipolar BBR shift of the energy level $|0\rangle$ can be expressed in terms of the
electric dipole transition matrix elements~\cite{FarWin81}
\begin{eqnarray}
\Delta E= -\frac{(\alpha T)^3}{2J_0+1} \sum_n |\langle 0||D||n \rangle|^2 F(y_n).
\label{dEg_F1}
\end{eqnarray}
Here $\alpha \approx 1/137$ is the fine-structure constant, $y_n \equiv (E_n - E_0)/T$, $T$ is the temperature, $J_0$
is the total angular momentum of the $|0\rangle$ state, $E_i$ is the energy of the $|i\rangle$ state, and $F(y)$ is the
function introduced by Farley and Wing~\cite{FarWin81}; its asymptotic expansion is given by
\begin{eqnarray}
F(y) \approx \frac{4\pi ^{3}}{45y}+%
\frac{32\pi ^{5}}{189y^{3}}+ \frac{32\pi ^{7}}{45y^{5}}+ \frac{512\pi ^{9}}{99y^{7}}.
\end{eqnarray}

Equation~(\ref{dEg_F1}) can be expressed in terms of the dc polarizability $\alpha_0$ of the $|0\rangle$ state as~\cite{PorDer06BBR},
\begin{equation}
\Delta E \equiv \Delta E^{\rm st} + \Delta E^{\rm dyn} ,
\label{delEg}
\end{equation}
where $\Delta E^{\rm st}$ and $\Delta E^{\rm dyn}$ are the static and dynamic parts, determined as
\begin{eqnarray}
\Delta E = -\frac{2}{15} (\alpha \pi)^3 T^4 \alpha_0\, [1 + \eta] .
\label{delEg_dyn}
\end{eqnarray}
Here $\eta$ represents a dynamic fractional correction to the total shift that reflects the averaging of the
frequency dependence of the polarizability over the frequency of the blackbody radiation spectrum.

The advantage of such a representation is a possibility to accurately measure the static part $\Delta E^{\rm st}$
and generally small contribution of the dynamic part. However, the recent measurement ~\cite{ArnKaeRoy18} of the
differential scalar dynamic polarizability of the  $^3\!D_1 -\, ^1\!S_0$ transition at  $\lambda = 10.6\, \mu{\rm m}$ yielded a very small value, $\Delta \alpha_0 (\lambda) =0.059(4)$ a.u.. An extrapolation to dc~\cite{ArnKaeRoy18} leads to even smaller value of the static scalar differential polarizability, $\Delta \alpha_0 (0) =0.018(6)$ a.u..
Therefore it is essential to evaluate the dynamic correction and its uncertainty.

The quantity $\eta$ can be approximated by~\cite{PorDer06BBR}
\begin{eqnarray}
&&\eta \approx \eta_1+\eta_2 \equiv \frac{80}{63\,(2J_0+1)} \, \frac{\pi^2}{\alpha_0 T} \nonumber \\
&&\times \sum_n \frac{ |\langle n||D||0 \rangle|^2}{y_n^3}
\left( 1+ \frac{21 \pi^2}{5\,y_n^2} \right) .
\label{eta}
\end{eqnarray}

Contributions of the intermediate odd-parity states to the dynamic fractional corrections $\eta_1$,
$\eta_2$, and $\eta = \eta_1 + \eta_2$ of the $6s^2\,^1\!S_0$, $5d6s\,^3\!D_{1,2}$, and $5d6s\,^1\!D_2$ states are presented in~\tref{table5}. Since the energy denominators in the first term of Eq.~(\ref{eta}) are proportional to $(E_n-E_0)^3$,
the sum over $n$ converges much more rapidly than for the polarizability [where the denominators are $\sim (E_n-E_0)$],
and the contribution of the states not listed in Table~\ref{table5} is expected to be negligible. Because the same matrix elements
are involved in the calculation of $\eta$ and the scalar polarizability $\alpha_0$ for a given state, we estimate that $\eta$ has
the same relative uncertainly as $\alpha_0$.

The corresponding static (dynamic) contributions to the BBR shift of a transition frequency are determined
by the differences of $\Delta E^{\rm st (dyn)}$ of the upper and lower states and in total
\begin{equation}
\Delta \nu = \Delta \nu^{\rm st} + \Delta \nu^{\rm dyn}.
\end{equation}
The static and dynamic BBR shifts for the $6s5d\;^{1,3}\!D_J -\, 6s^2\,^1\!S_0$ transitions at $T=300$~K are given in Table~\ref{table5a}.

The theoretical values of the $^3\!D_{1,2}$ and $^1\!S_0$ polarizabilities are very
close to each other. Taking into account the theoretical uncertainties, we are unable to predict reliably  the
differential polarizabilities and $\Delta \nu^{\rm st}$ for these transitions.
For this reason the values of $\Delta \nu^{\rm st}$ for the
$^3\!D_{1,2} -\, ^1\!S_0$ transitions, presented in Table~\ref{table5a}, are found using the experimental results for $\Delta \alpha_0 (^3\!D_{1,2} -\, ^1\!S_0)$~\cite{ArnKaeRoy18}.
The polarizabilities of the $^1\!S_0$ and $^1\!D_2$ states differ more significantly and we obtain
$\Delta \nu^{\rm st}=66(17)$~mHz  for the $^1\!D_2 -\, ^1\!S_0$ transition.

In contrast with the scalar static polarizabilities, the $\eta$ corrections for the ground and $^3\!D_J$ states differ by
a factor of two and we estimate the uncertainties of our values of $\Delta \nu^{\rm dyn}$ for the $^{3,1}\!D_J -\, ^1\!S_0$ transitions to be 12-18\%.

We would like to emphasise that  the $^3\!D_1 -\, ^1\!S_0$ transition is unique in the sense that the static BBR frequency shift
is two times (in absolute value) smaller than the dynamic BBR shift. The total BBR frequency shift
for this transition is very small, $\Delta \nu = -0.48$ mHz. This value is in excellent agreement with the result
obtained in Ref.~\cite{ArnKaeRoy18}.

Since the differential scalar static polarizability of the  $^3\!D_1 -\, ^1\!S_0$ transition is close to zero,
we have also considered the third-order contribution to this
quantity, involving two interactions of the electric-dipole operator $\bf D$
with the external electric field and one hyperfine interaction~\cite{AngSan68} (see also Ref.~\cite{BelSafDer06}
for further details). We estimated this contribution to be $10^{-4}-10^{-5}$ a.u., resulting in the BBR frequency shift below 1 $\mu$Hz, negligible at the present level of accuracy.

\begin{table} [t]
\caption{\label{table5a}
The dynamic corrections $\Delta E^{\rm dyn}/h$ ($h$ is the Planck constant) and $\Delta \nu^{\rm st\, (\rm dyn)}$ to the BBR shifts for the $5d6s\;^{3,1}\!D_J$ and $6s^2\;^1\!S_0$ states and the $^{3,1}\!D_J -\, ^1\!S_0$ transitions, respectively, at $T=300$~K.
Static scalar polarizabilities $\alpha_0$ are listed. The uncertainties are given in parentheses.}
\begin{ruledtabular}
\begin{tabular}{lccc}
&\multicolumn{1}{c}{$\alpha_0$ (a.u.)} & \multicolumn{1}{c}{$\eta$} & \multicolumn{1}{c}{$\Delta E^{\rm dyn}/h$ (mHz)} \\
\hline     \\[-0.6pc]
$6s^2~^1\!S_0$ & 63.0(0.8)$^{\rm a}$& 0.00051(1)&	-0.277(5)  \\
$5d6s~^3\!D_1$ & 63.5(2.8)$^{\rm a}$& 0.00110(5)&	-0.603(38) \\
$5d6s~^3\!D_2$ & 62.1(2.6)$^{\rm a}$& 0.00092(4)&	-0.494(29) \\
$5d6s~^1\!D_2$ & 55.3(1.7)          & 0.00081(2)&	-0.387(17)\\ [0.2pc]
\hline     \\[-0.7pc]
&\multicolumn{1}{c}{$\Delta \nu^{\rm st}$ (mHz)} & \multicolumn{1}{c}{} & \multicolumn{1}{c}{$\Delta \nu^{\rm dyn}$ (mHz)} \\
\hline     \\[-0.6pc]
$^3\!D_1 -\, ^1\!S_0$ & -0.15(5)$^{\rm b}$ &            &       -0.33(4)\\
$^3\!D_2 -\, ^1\!S_0$ &  10.1(8)$^{\rm b}$ &            &       -0.22(3)\\
$^1\!D_2 -\, ^1\!S_0$ &  66(17)            &            &       -0.11(2)\\
\end{tabular}
\end{ruledtabular}
\flushleft
$^{\rm a}$These values are taken from~\cite{PaeArnHaj16}; \\
$^{\rm b}$Extracted from the experimental results~\cite{ArnKaeRoy18}.
\end{table}
\section{Fine-structure constant variation}
\label{alphas}
Since frequencies of atomic clocks have different dependencies on the fine-structure constant $\alpha$, one can search for
the $\alpha$-variation by precisely measuring ratios of two clocks frequencies over time \cite{SafBudDem18}.
This subject recently became of  even higher interest, since
the variation of the fundamental constants was directly
linked to the dark matter searches \cite{Arv15,RobBleDai17,KalYu17}.

To evaluate the sensitivity of the particular clock to the variation of $\alpha$, one
calculates the relativistic frequencies shifts, determined by so-called $q$ factors, according to
\begin{equation}
\omega(x) = \omega'+ qx,
\label{omega}
\end{equation}
where $\omega'$ is the present laboratory value of the frequency, $x = (\alpha/\alpha')^2 - 1$,
and the $q$ factor is determined as
\begin{equation}
q = \left. \frac{d\omega}{dx}\right|_{x=0} .
\end{equation}

From \eref{omega} we can easily obtain
\begin{equation}
\frac{\Delta \omega}{\omega} \approx Q\, \frac{\Delta \alpha}{\alpha},
\label{Del_om}
\end{equation}
where $Q \equiv 2q/\omega$, $\Delta \omega \equiv \omega - \omega'$, and $\Delta \alpha \equiv \alpha - \alpha'$.
\begin{table}[htb]
\caption{\label{qQ} The $q$ and $Q$ factors for the $^{3,1}\!D_J -\, ^1\!S_0$ transitions
are obtained in the CI, CI+MBPT, and CI+all-order approximations.}
\begin{ruledtabular}
\begin{tabular}{llccc}
\multicolumn{1}{c}{} & \multicolumn{1}{c}{} & \multicolumn{1}{c}{$q$}  & \multicolumn{1}{c}{$Q$} \\
\hline \\ [-0.3pc]
   CI   &  $^1\!S_0$ &    ---    &        \\[0.1pc]
        &  $^3\!D_1$ &   14380   &  2.44   \\[0.1pc]
        &  $^3\!D_2$ &   14572   &  2.34   \\[0.1pc]
        &  $^3\!D_3$ &   15257   &  2.15  \\[0.1pc]
        &  $^1\!D_2$ &   16515   &  1.91 \\ [0.3pc]

CI+MBPT &  $^1\!S_0$ &    ---    &        \\[0.1pc]
        &  $^3\!D_1$ &   14951   &  2.54   \\[0.1pc]
        &  $^3\!D_2$ &   15437   &  2.48   \\[0.1pc]
        &  $^3\!D_3$ &   17223   &  2.43  \\[0.1pc]
        &  $^1\!D_2$ &   19061   &  2.20 \\ [0.3pc]

CI+All  &  $^1\!S_0$ &   ---     &        \\[0.1pc]
        &  $^3\!D_1$ &   14854   &  2.52   \\[0.1pc]
        &  $^3\!D_2$ &   15294   &  2.46   \\[0.1pc]
        &  $^3\!D_3$ &   16873   &  2.38  \\[0.1pc]
        &  $^1\!D_2$ &   18633   &  2.15
\end{tabular}
\end{ruledtabular}
\end{table}
\subsection {A simple estimate}
In a single-electron approximation the relativistic energy shift is given in Refs.~\cite{DzuFlam08b,DzuFlaWeb99}.
For a singly-charged ion it can be rewritten (in a.u.) as
\begin{equation}
\Delta_a = -\sqrt{\frac{|\varepsilon_a|^3}{2}}\, (\alpha Z)^2 \left[ \frac{1}{j_a+1/2} - C(Z,j_a,l_a) \right] ,
\label{Dela}
\end{equation}
where $a$ is the index for a single-electron state, $\varepsilon_a$ is its energy ($\varepsilon_a = -|\varepsilon_a|$),
$j_a$ and $l_a$ are the total and orbital angular momenta of the state $a$, and $C(Z,j_a,l_a)$
is a parameter introduced to simulate the effect of the Hartree-Fock exchange interaction
and other many-body effects. An accurate value of $C(Z,j_a,l_a)$ can be obtained only from many-body calculations
but $C(Z,j_a,l_a) \approx 0.6$~\cite{DzuFlaWeb99} can be used for a rough estimate.

If we approximate the transition from the upper to lower state by a single-electron $b - a$ transition,
the $q$ factor can be written as
\begin{equation}
q \approx \Delta_b - \Delta_a.
\label{q}
\end{equation}

Using \eref{q} we are able to roughly estimate the $q$ (and $Q$) factors for the
$^{3,1}\!D_J -\, ^1\!S_0$ transitions. Taking into account that the main relativistic configurations are
$6s^2$ for $^1\!S_0$, $6s5d_{3/2}$ for $^3\!D_{1,2}$, and $6s5d_{5/2}$ for $^1\!D_2$, we can approximate
the $^3\!D_{1,2} -\, ^1\!S_0$ transitions by the single-electron $5d_{3/2} - 6s$ transition and
the $^1\!D_2 -\, ^1\!S_0$ transition by the single-electron $5d_{5/2} - 6s$ transition.

Taking into account that \eref{Dela} is valid for an external electron above closed shells, for an estimate, we substitute the Hartree-Fock energies of the univalent Lu$^{2+}$
$\varepsilon_{6s} \approx -0.73$, $\varepsilon_{5d_{3/2}} \approx -0.70$, and $\varepsilon_{5d_{5/2}} \approx -0.69$ a.u.
to \eref{Dela}.

Then we obtain $\Delta_{6s} \approx -0.047$, $\Delta_{5d_{3/2}} \approx 0.011$, and $\Delta_{5d_{5/2}} \approx 0.029$ a.u..
It gives us the following transition $q$ factors:
\begin{eqnarray*}
q (^3\!D_{1,2} -\, ^1\!S_0) \simeq \Delta_{5d_{3/2}} - \Delta_{6s} &\approx& 0.058 \text{ a.u.} \\
                                                                   &\approx& 12700 \text{ cm}^{-1} \\
q (^1\!D_2 -\, ^1\!S_0)     \simeq \Delta_{5d_{5/2}} - \Delta_{6s} &\approx& 0.076 \text{ a.u.} \\
                                                                   &\approx& 16700 \text{ cm}^{-1} .
\end{eqnarray*}
Taking into account that $\omega(^3\!D_{1,2} -\, ^1\!S_0) \simeq 12100 \text{ cm}^{-1}$ and
$\omega( ^1\!D_2 -\, ^1\!S_0) = 17333 \text{ cm}^{-1}$ we obtain the following estimate:
\begin{eqnarray*}
Q (^{3,1}\!D_J -\, ^1\!S_0) \approx 2 .
\end{eqnarray*}
\subsection {Full-scale calculation}
We carried out calculations in three approximations: CI, CI+MBPT, and CI+all-order. In each case three calculations
(with $x=0,\pm1/8$) were done. Using Eqs.~(\ref{omega})-(\ref{Del_om}), we found the $q$ and $Q$ factors
for the transitions from the $^{3,1}\!D_J$ states to the ground state. The results are presented in~\tref{qQ}.

The results obtained from the simple estimate above are in good agreement with  those obtained from the full-scale calculation. Thus, if a high accuracy calculation of the $q$ factors
is not needed, Eq.~(\ref{Dela}) can be used for a quick estimate of these quantities.

The factors $Q \approx 2-2.5$ presented in \tref{qQ}, though smaller than $|Q|=15$ found recently for an optical transition in Yb~\cite{SafPorSan18}, are larger than the values for all currently operating clocks with the
exception of the Hg$^+$ and octupole Yb$^+$ clocks \cite{FlaDzu09}. Thus, the clocks based on the $6s^2\,^1\!S_0 -\, 5d6s\,^3\!D_{1,2}$ transitions in Lu$^+$ are good candidates for the search for the $\alpha$-variation.

\section{Conclusion}
\label{Concl}
To conclude, we carried out the calculations of the quadrupole moments and the corresponding quadrupole shifts, demonstrating the need to accurately suppress these effects. We provided the recommended value of the dc $5d6s\,^1\!D_2$ polarizability
and established the dominant contributions of the intermediate states to the polarizability.
We determined the dynamic BBR shifts for the $6s^2\,^1\!S_0$, $5d6s\,^3\!D_{1,2}$ and $5d6s\,^1\!D_2$ energy levels.
The values of the dynamic BBR shifts at $T=300$~K for the clock transitions are determined with 12-18\%  uncertainties.

We note that the differential polarizability of the $6s^2\,^1\!S_0 -\, 5d6s\,^1\!D_2$ transition is negative (so the micromotion
effect can be canceled at the magic radio-frequency). The $5d6s \,^1\!D_2$ state has very small quadrupole moment which may eliminate
a need for hyperfine averaging. These features make the $6s^2\,^1\!S_0 -\, 5d6s\,^1\!D_2$ transition a good
candidate for creating a clock in its own right.

We confirm the observation of Ref.~\cite{ArnKaeRoy18} that the Lu$^+$ $6s^2\,^1\!S_0 - 5d6s\,^3\!D_{1}$ transition is the only known clock transition where the dynamic part of the BBR frequency shift is much larger than the static part. We also considered the third-order contribution to this differential polarizability and estimated that the resulting BBR frequency shift is negligible.
Finally, we calculated the sensitivity of the Lu$^+$ clock transitions to the variation of the fine-structure constant and related dark matter searches.

\section*{Acknowledgement}
We thank Murray Barrett for helpful discussions, comments on the manuscript and bringing our attention to the possible
use of the $^1\!D_2$ state for diagnostic and the $^1\!S_0 -\, ^1\!D_2$ transition as the potential clock transition. We are also grateful to Vladimir Dzuba for bringing our attention to issues with calculating the $q$ and $Q$ factors and useful remarks.
This research was performed in part under the sponsorship of the Office of Naval Research, award number N00014-17-1-2252. S.~P. acknowledges support from Russian Foundation for Basic Research under Grant No. 17-02-00216.


\end{document}